\documentclass[aps,prb,reprint,superscriptaddress,longbibliography]{revtex4-2}

\usepackage{amsmath,amssymb,bm}
\usepackage{graphicx}
\usepackage[dvipdfmx]{hyperref}
\usepackage{physics}
\usepackage{color}
\usepackage{comment}

\begin{document}

\title{Topological-Mass Control of an Emergent Kondo Scale in an Interacting SSH Chain}

\author{Ryosuke Yoshii}
\affiliation{Center for Liberal Arts and Sciences, Sanyo-Onoda City University, Onoda Yamaguchi 756-0884, Japan}

\author{Rio Oto}
\affiliation{Department of Engineering, Sanyo-Onoda City University, Onoda, Yamaguchi 756-0884, Japan}

\date{\today}

\begin{abstract}
Topological bound states emerging at domain walls of dimerized chains provide a robust platform for exploring correlation effects beyond single-particle physics. 
When such a soliton state is coupled to a metallic substrate, local Coulomb interactions can give rise to Kondo screening. Here we demonstrate analytically and numerically that, in an interacting Su–Schrieffer–Heeger (SSH) chain, the Kondo temperature is directly controlled by the topological mass that governs the bulk gap. 
Near the topological transition, the Kondo scale collapses linearly with the mass parameter while retaining its exponential sensitivity to hybridization. 
This establishes a minimal mechanism by which a bulk topological parameter quantitatively determines an emergent many-body energy scale. 
Our results clarify the strong configuration dependence of soliton-induced Kondo signatures observed in graphene nanoribbon systems on Au(111) and provide experimentally testable predictions for scanning tunneling spectroscopy.
\end{abstract}

\maketitle

% ============================================================
\section{Introduction}
% ============================================================

A central question in condensed matter physics is whether topology can directly regulate many-body energy scales. While topological electronic states localized at defects and boundaries provide robust platforms for novel quantum phenomena \cite{Su1979, Heeger1988}, their influence on correlation-driven screening processes remains largely unexplored. In one-dimensional systems, dimerized molecular chains realizing the Su-Schrieffer-Heeger (SSH) model host midgap soliton states at domain walls separating distinct topological phases. These localized states offer a natural setting to investigate how topological mass terms may control many-body screening phenomena.

Recent advances in scanning tunneling microscopy (STM) have enabled the observation of such topological states in molecular chains on metallic substrates like Au(111). Thanks to these advances, experiments have successfully demonstrated the engineering of topological phases and the observation of localized soliton states in graphene nanoribbons \cite{Groning2018, Rizzo2018}. While these studies establish the existence of robust topological midgap states, the interplay between such localized orbitals and many-body correlations—specifically the emergence and energy scale of Kondo screening—remains to be fully explored. A particularly intriguing frontier is the intersection of topology and strong correlations \cite{Rachel2018}, namely whether a localized spin-1/2 moment formed in a soliton state can be screened by the substrate's conduction electrons and how the associated Kondo temperature is governed by the underlying topological structure \cite{Kondo1964, Hewson}. While the theoretical possibility of Kondo screening in SSH-type systems has been discussed in recent literature \cite{Aligia2025}, several critical issues regarding experimental realization remain unresolved. One of the key challenges is the role of Coulomb repulsion: without sufficient on-site interaction, double occupancy is not suppressed and a stable local moment cannot form. A recent report on the Class-II OInIn isomer suggests that both an SSH-like band structure and strong Coulomb repulsion may coexist in a realistic system \cite{Ortiz2025}, providing a promising platform to investigate topology-controlled many-body screening.

One of the most significant challenges is the observed heterogeneity in experimental signatures. 
Even in atomically precise zigzag graphene nanoribbons, the observed topological signatures can be sensitive to the local environment and structural defects \cite{Ruffieux2016}. 
This inherent heterogeneity in experimental spectra suggests that the many-body physics, such as the Kondo effect, may emerge selectively depending on the local coupling to the substrate. 
Indeed, even within the same molecular chain, Kondo resonances often appear clearly at some sites while being entirely absent at others \cite{Li2020}. 
Such environmental sensitivity has been classically demonstrated in molecular systems like cobalt phthalocyanine on Au(111), where the Kondo effect can be "switched" or dramatically tuned by modifying the chemical bonding with the substrate \cite{Zhao2005}. 
While this highlights the critical role of adsorption geometry for magnetic ions, it remains an open question how this sensitivity manifests in states inherently protected by band topology.
While previous studies have focused on the universal aspects of the topological Kondo effect, a systematic understanding of this experimental variability remains elusive. 
Furthermore, a direct analytical link between the spatial properties of the topological wave function and the resulting spectroscopic line shapes is needed to reliably identify the Kondo origin of zero-bias anomalies.

In this paper, we demonstrate that topology can directly control a many-body energy scale by investigating Kondo screening at domain-wall solitons on metallic substrates like Au(111). 
Starting from the SSH-Hubbard framework, we analytically derive an effective Anderson impurity model that captures both the renormalized interaction and hybridization parameters of the soliton state. 
We show that the Kondo temperature $T_{\rm K}$ is governed by the topological dimerization parameter and vanishes linearly at the topological transition, establishing a direct proportionality between topological mass and many-body screening scale. In addition, $T_{\rm K}$ remains exponentially sensitive to the hybridization strength $\Gamma$, which depends strongly on adsorption geometry. 
This is in agreement with the high sensitivity observed in experiments \cite{Franke2011}.  
This combined topological and geometrical control naturally explains the order-of-magnitude variations in $T_{\rm K}$ observed experimentally. Finally, we provide quantitative predictions for scanning tunneling spectroscopy, including Fano interference and its temperature evolution.

The paper is organized as follows. 
In Sec.~II, we introduce the Su-Schrieffer-Heeger-Hubbard (SSH-Hubbard) model for the Class-II oligo(indenoindene) chain and discuss the effective Coulomb interaction $U_{\rm eff}$ renormalized by the metallic substrate. 
Sec.~III describes the mapping of the localized soliton state onto an effective Anderson impurity model to investigate the many-body physics. 
In Sec.~IV, we analyze the exponential sensitivity of the Kondo temperature $T_{\rm K}$ to the adsorption geometry and the resulting heterogeneity in experimental observations. 
Sec.~V provides theoretical predictions for scanning tunneling spectroscopy (STS), incorporating the Fano interference between the soliton resonance and the Au(111) continuum. 
Finally, we summarize our findings and discuss future outlooks in Sec. VI.

% ============================================================
\section{Model: Dimerized chain and soliton orbital}
\label{Model}
% ============================================================

\begin{figure}
\centering
\includegraphics[clip,width=9cm]{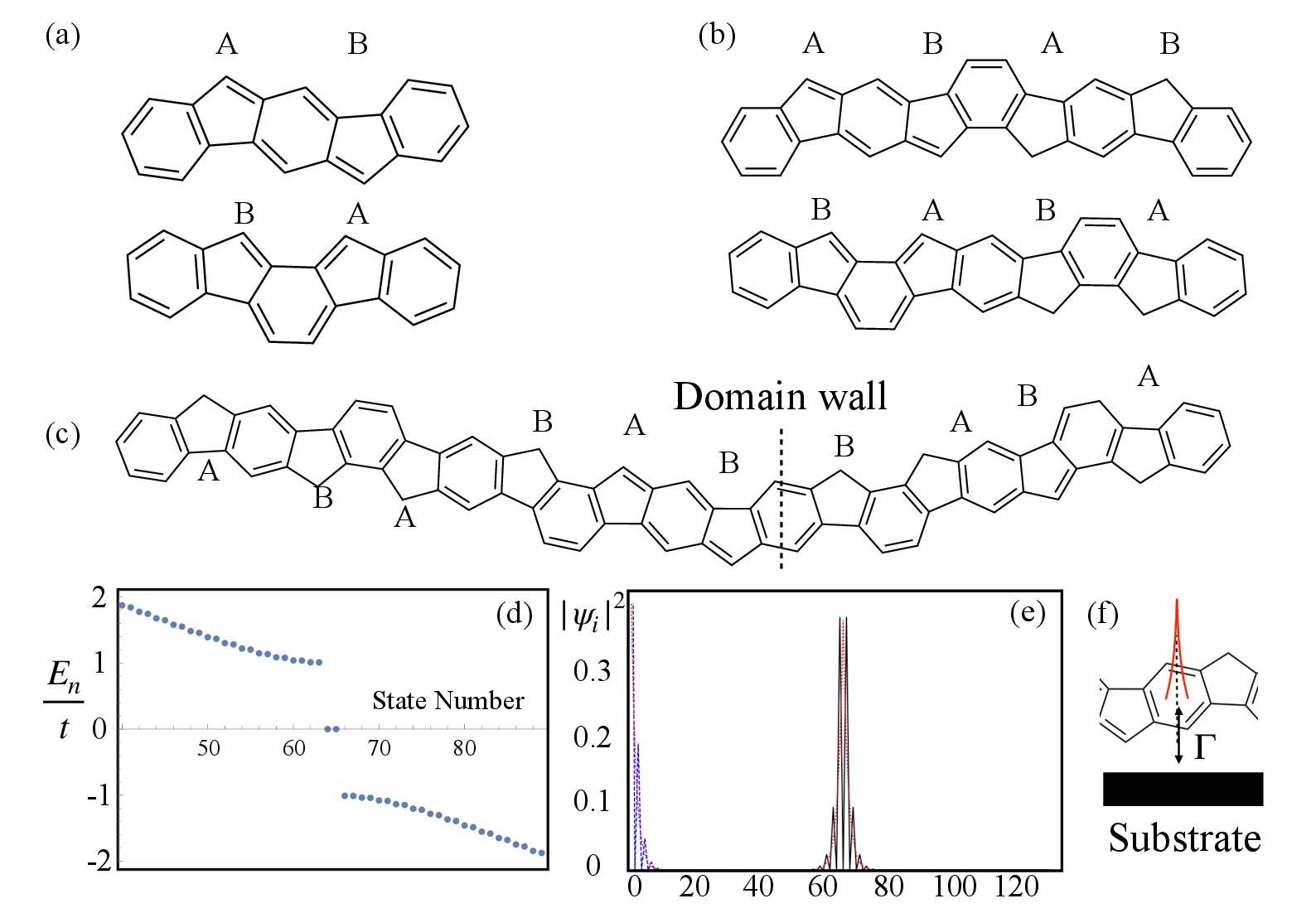}
\caption{
Chemical structures of the system in this work. 
(a) Molecular structure of Class-II OInIn isomers with alternating hopping amplitudes. 
(b) Schematic of a long chain formed by connecting these units. 
(c) Formation of a domain wall at the boundary between two distinct dimerization patterns. 
(d) Calculated energy spectrum of the SSH model ($t_1=2$, $t_2=1$) with 2N=128 sites, showing a clear midgap state at E=0 due to the topological defect.  
(e) Spatial distribution of the probability density $|\psi_i|^2$ for the soliton state (solid line) localized at the domain wall and the edge state (dashed line). 
Dotted lines represent the approximated solutions derived in the later section. 
(f) Conceptual illustration of the soliton orbital hybridized with the Au(111) substrate (modeled as a conduction bath), forming the basis for the effective Anderson impurity model.
}
\label{fig:ChemStructure}
\end{figure}

We consider a one-dimensional dimerized chain described by a tight-binding Hamiltonian of SSH type which describes the Class-II OInIn isomers in Ref.~\cite{Ortiz2025} depicted in Fig~\ref{fig:ChemStructure}, 
\begin{equation}
H_{\rm SSH}=
\sum_{n=1}^N
\sum_{\sigma=\uparrow,\downarrow}
\left[
t_1 c^\dagger_{n,A,\sigma}c_{n,B,\sigma}
+
t_2 c^\dagger_{n,B,\sigma}c_{n+1,A,\sigma}
+{\rm h.c.}
\right],
\label{eq:SSH}
\end{equation}
where $c_{n,A,\sigma}$ ($c^\dagger_{n,A,\sigma}$) and $c_{n,B,\sigma}$ ($c^\dagger_{n,B,\sigma}$), respectively, represent the annhilation (creation) operator of electron with spin $\sigma$ at $A$ part of site $n$ and $B$ part of site $n$. 
Here,  $t_1$ and $t_2$ denote alternating hopping amplitudes, and the total number of sites is set to be $2N$. 
The chain supports two topologically distinct dimerization patterns. 
A domain wall between them generates a soliton-bound state localized near the topological defect. 

At the noninteracting level, the soliton state produces a midgap resonance whose wave function decays exponentially away from the domain wall. 
In Fig.~\ref{fig:ChemStructure} we show the domain wall structure, energy spectrum, and localized modes. 
The domain wall appears at the boundary of the two structures in Fig.~(b) (Fig.~(c)). 
The energy spectrum for the SSH model with $t_1=2$ and $t_2=1$ is plotted in Fig.~(d).  
The total number of sites is set to $128$, and the topological defect is located between the site numbers $64$ and $65$ counted from the far left. 
The solid and dashed lines in Fig.~(e) and Fig.~(f), respectively, represent the spatial configurations of the electron probability density $|\psi_i|^2$ for the solitonic state localized in the vicinity of the topological defect and for the edge state, where $\psi_i$ represents the amplitude of the wave function at site $i$.  
Denoting the soliton orbital by operator $d_\sigma$, we represent the low-energy physics in terms of an effective localized orbital with energy $\epsilon_d$.

To incorporate correlation effects, we include an effective on-site repulsion on the soliton orbital,
\begin{equation}
H_{\rm int}=U_{\rm eff} n_{d\uparrow}n_{d\downarrow},
\label{eq:Hint}
\end{equation}
where $n_{d\sigma}=d^\dagger_\sigma d_\sigma$. 
The soliton orbital thus behaves as a correlated impurity embedded in the chain.

The effective on-site Coulomb interaction, $U_{\rm eff}$, is significantly renormalized from its bare value due to the screening effects of the metallic Au(111) substrate. 
When a molecule is adsorbed on a metal surface, the image-charge effect typically reduces the electron-electron repulsion from several electron volts in the gas phase to a much smaller effective value \cite{Neaton2006}. 
In evaluating the effective on-site Coulomb interaction $U_{\rm eff}$, we refer to the energy scales reported in spectroscopic studies of topological nanostructures \cite{Groning2018, Rizzo2018}. Due to the screening effects from the metallic Au(111) substrate \cite{Neaton2006}, the Coulomb penalty is significantly reduced, justifying our choice of $U_{\rm eff}=0.2-0.5$ eV for describing the correlated soliton state.

The effective coulomb potential strength $U_{\mathrm{eff}}$ can be evaluated as follows. 
Let us call the wave function of the solitonic mode $\psi^{\rm soliton}$ and its amplitude at site $i$ to be $\psi^{\rm soliton}_i$. 
The Coulomb interaction term is projected into the solitonic mode basis as 
\begin{equation}
U\sum_i |\psi_i^{\rm soliton}|^4 n_{d\uparrow}n_{d\downarrow}.
\end{equation}
Thus, the effective Coulomb interaction becomes 
\begin{equation}
U_{\rm eff}=U\sum_i |\psi_i^{\rm soliton}|^4. 
\label{eq:Ueff}
\end{equation}
It is noteworthy that the presence of $|\psi_i^{\rm soliton}|^4$ highly reflects the topological phase transition, namely, it becomes zero if the mid gap state vanishes ($t_1=t_2$). 
For later convenience, we assume the topological defect is located between $N_{\rm s}$ site and $N_{\rm s}+1$ site. 
If the localization is tight, one can further evaluate the effective Coulomb interaction as 
\begin{equation}
U_{\rm eff}\simeq U|\psi_{N_{\rm s}}^{\rm soliton}|^4+U|\psi_{N_{\rm s}+1}^{\rm soliton}|^4. 
\label{eq:Ueff2}
\end{equation}

% ============================================================
\section{Coupling to Au(111): Effective Anderson impurity model}
% ============================================================

When the chain is adsorbed on Au(111), electrons may tunnel between the soliton orbital and the metallic substrate. 
The substrate is modeled as a conduction bath
\begin{equation}
H_{\rm bath}=
\sum_{k,\sigma}\epsilon_k c^\dagger_{k\sigma}c_{k\sigma}.
\label{eq:Hbath}
\end{equation}
Hybridization between the soliton orbital and substrate is described by
\begin{equation}
H_{\rm hyb}=
\sum_{k,\sigma}
\left(
V_k c^\dagger_{k\sigma} d_\sigma
+
{\rm h.c.}
\right).
\label{eq:Hhyb}
\end{equation}
The full effective model is therefore the single-impurity Anderson Hamiltonian
\begin{equation}
H=H_{\rm bath}
+\sum_\sigma \epsilon_d d^\dagger_\sigma d_\sigma
+U_{\rm eff}n_{d\uparrow}n_{d\downarrow}
+H_{\rm hyb}.
\label{eq:Anderson}
\end{equation}

The hybridization strength is characterized by
\begin{equation}
\Gamma(\omega)=\pi\sum_k |V_k|^2\delta(\omega-\epsilon_k).
\label{eq:Gammaomega}
\end{equation}
For Au(111) in the relevant energy window, we approximate $\Gamma(\omega)\approx \Gamma$ as constant.
This hybridization strength $\Gamma$ is approximately evaluated from the solitonic mode as the case of $U_{\rm eff}$. 
Since the solitonic mode has a significant spatial modulation of the density of electrons, the hybridization becomes 
\begin{equation}
\Gamma_i=\Gamma_0 |\psi^{\rm soliton}_i|^2,  
\label{eq:Gammai}
\end{equation}
where $\Gamma_0$ is a constant determined by the density of state in the substrate and the hopping between the chain and the substrate. 
This $\Gamma_i$ becomes zero except near the topological defect and furthermore, it vanishes in the non-topological phase.

The physical interpretation is straightforward: the topological domain wall generates a localized orbital; Coulomb repulsion stabilizes a local moment; and hybridization with the Au substrate enables Kondo screening.

% ============================================================
\section{Topological parameter and effective Coulomb repulsion} 
% ============================================================

In the Sec.~\ref{Model}, we have shown the rough estimation of $U_{\rm eff}$. 
In this part, we further evaluate the effective Coulomb interaction by using the approximate solution of the solitonic mode in the SSH model. 
The solitonic mode near the domain wall is given by (See Fig.~\ref{fig:ChemStructure} e)
\begin{equation}
\psi_i^{\rm soliton}\approx Ar^{|i-N_{\rm s}|}, 
\end{equation}
where we introduce the normalization factor $A$ and the parameter $r$ as 
\begin{equation}
r:=\sqrt{\left|\frac{t_1}{t_2}\right|}.  
\end{equation}
Here we assume $|t_1|<|t_2|$. 
This approximated solution yields the localization length $\xi$ as a function of $r$ as 
\begin{equation}
\xi =-\frac{1}{\ln r}. 
\label{eq:xiandr}
\end{equation}
Here we note that the physical localization length $\xi_{\rm phys} $ must be defined as $\xi_{\rm phys}=-a/{\ln r}$ with a lattice constant $a$, so that the solitonic mode becomes 
\begin{equation}
\psi_i^{\rm soliton}\propto  e^{-a|i-N_{\rm s}|/\xi}. 
\end{equation}

The nomalization of the solitonic mode $\sum_i|\psi_i^{\rm soliton}|^2=1$ yields 
\begin{equation}
\sum_i|\psi_i^{\rm soliton}|^2=A^2 \sum_{m=0}^{\infty} 2r^{2m} =2A^2 \frac{1}{1-r^2}=1, 
\end{equation}
where we assume the domain wall is far from the edge and the solitonic mode becomes rapidly zero so that we can change the upper bound of the summation to be $\infty$. 
Thus we obtain 
\begin{equation}
|\psi_i^{\rm soliton}|^2=\frac{1-r^2}{2} r^{2|i-N_{\rm s}|}.  
\label{eq:solitonapprox}
\end{equation}
By using this approximated solution, we can evvaluate $U_{\rm eff}$ as follows, 
\begin{align}
U_{\rm eff}
&=U\sum_i |\psi_i^{\rm soliton}|^4=U\frac{(1-r^2)^2}{4} \cdot 2\sum_{m=0}^{\infty} r^{4m}\nonumber\\
&=\frac{1}{2}U\frac{(1-r^2)^2}{1-r^4}=\frac{1}{2}U\frac{1-r^2}{1+r^2}.  
\label{eq:UeffTopPara}
\end{align}
This expression shows the direct connection between the parameter determining the topological soliton and the effective Coulomb interaction in the solitonic mode. 
Eq.~\eqref{eq:UeffTopPara} can also be written by using the localization length \eqref{eq:xiandr} which may be better quantity for experiments as 
\begin{align}
U_{\rm eff}
&=\frac{1}{2}U\frac{1-e^{-2/\xi}}{1+e^{-2/\xi}}=\frac{1}{2}U\tanh \frac{1}{\xi}.   
\label{eq:UeffTopxi}
\end{align}

Similarly, the hybridization \eqref{eq:Gammai} can also evaluated by using \eqref{eq:solitonapprox} as
\begin{equation}
\Gamma_i=\frac{1-r^2}{2}\Gamma_0 r^{2|i-N_{\rm s}|}.   
\label{eq:GammaiTop}
\end{equation}
This expression implies the power-law decrease of the hybridization away from the domain wall.

% ============================================================
\section{Kondo regime and Schrieffer--Wolff reduction}
% ============================================================

In the local-moment regime, $\epsilon_d<0$ and $\epsilon_d+U_{\rm eff}>0$, the soliton orbital is singly occupied and behaves as a spin-$1/2$ impurity. 
Performing a Schrieffer--Wolff transformation yields an effective Kondo model \cite{Hewson}, 
\begin{equation}
H_{\rm Kondo}=H_{\rm bath}
+J\bm{S}\cdot\bm{s}(0),
\label{eq:Kondo}
\end{equation}
where $\bm{S}$ is the impurity spin and $\bm{s}(0)$ is the conduction-electron spin density at the impurity site.

The antiferromagnetic exchange coupling is
\begin{equation}
J = 2|V|^2
\left(
\frac{1}{|\epsilon_d|}
+
\frac{1}{\epsilon_d+U_{\rm eff}}
\right).
\label{eq:J_SW}
\end{equation}
Again, this $J$ has a significant spatial dependence stemming from $|V|\propto |\psi_i^{\rm soliton}|^2$. 
The Kondo temperature is then approximately
\begin{equation}
T_{\rm K} \sim D \exp\left(-\frac{1}{\rho J}\right),
\label{eq:TK_def}
\end{equation}
where $D$ is a bandwidth cutoff and $\rho$ is the substrate density of states.

At particle-hole symmetry $\epsilon_d=-U_{\rm eff}/2$, one obtains a widely used approximation,
\begin{equation}
T_{\rm K} \sim \sqrt{\Gamma_{N_{\rm s}} U_{\rm eff}}
\exp\left(-\frac{\pi U_{\rm eff}}{8\Gamma_{N_{\rm s}}}\right),
\label{eq:TK_PH}
\end{equation}
which will be used below to quantify experimental scales. 
Here, we assume the localization of the zero mode is tight enough to approximate as a one-site peak at the defect site. 
As mentioned before, the two parameters $\Gamma$ and $U_{\rm eff}$ reflect the soliton shape via Eqs.~\eqref{eq:Ueff} and \eqref{eq:Gammai}. Moreover, both vanish if the system is in the non-topological sector.

% ============================================================
\section{Scanning tunneling spectroscopy and Fano interference}
% ============================================================

The differential conductance $dI/dV$ measured in STS can be described by the Fano line shape, which arises from the quantum interference between the discrete Kondo resonance and the substrate's conduction continuum \cite{Madhavan1998}. 
The profile is characterized by the Fano parameter $q$, where $q\to \infty$ yields a Lorentzian peak and $q=0$ results in a symmetric dip \cite{Fano1961}.

The standard expression is
\begin{equation}
\frac{dI}{dV}(V)
=
a\frac{(q+\epsilon)^2}{1+\epsilon^2}+b,
\qquad
\epsilon=\frac{eV-V_0}{\Gamma_{\rm K}},
\label{eq:Fano}
\end{equation}
where $q$ is the Fano parameter, $V_0$ is the resonance center, and $\Gamma_{\rm K}$ is the resonance width.

For Kondo physics, $\Gamma_{\rm K}$ is controlled by $T_{\rm K}$ and satisfies $k_BT_{\rm K}\sim \Gamma_{\rm K}$. 
Thus, fitting STS spectra to Eq.~(\ref{eq:Fano}) provides a direct experimental estimate of $T_{\rm K}$.

% ============================================================
\section{Results}
% ============================================================

\subsection{Exponential sensitivity of the Kondo temperature}

\begin{figure}
\centering
\includegraphics[clip,width=8cm]{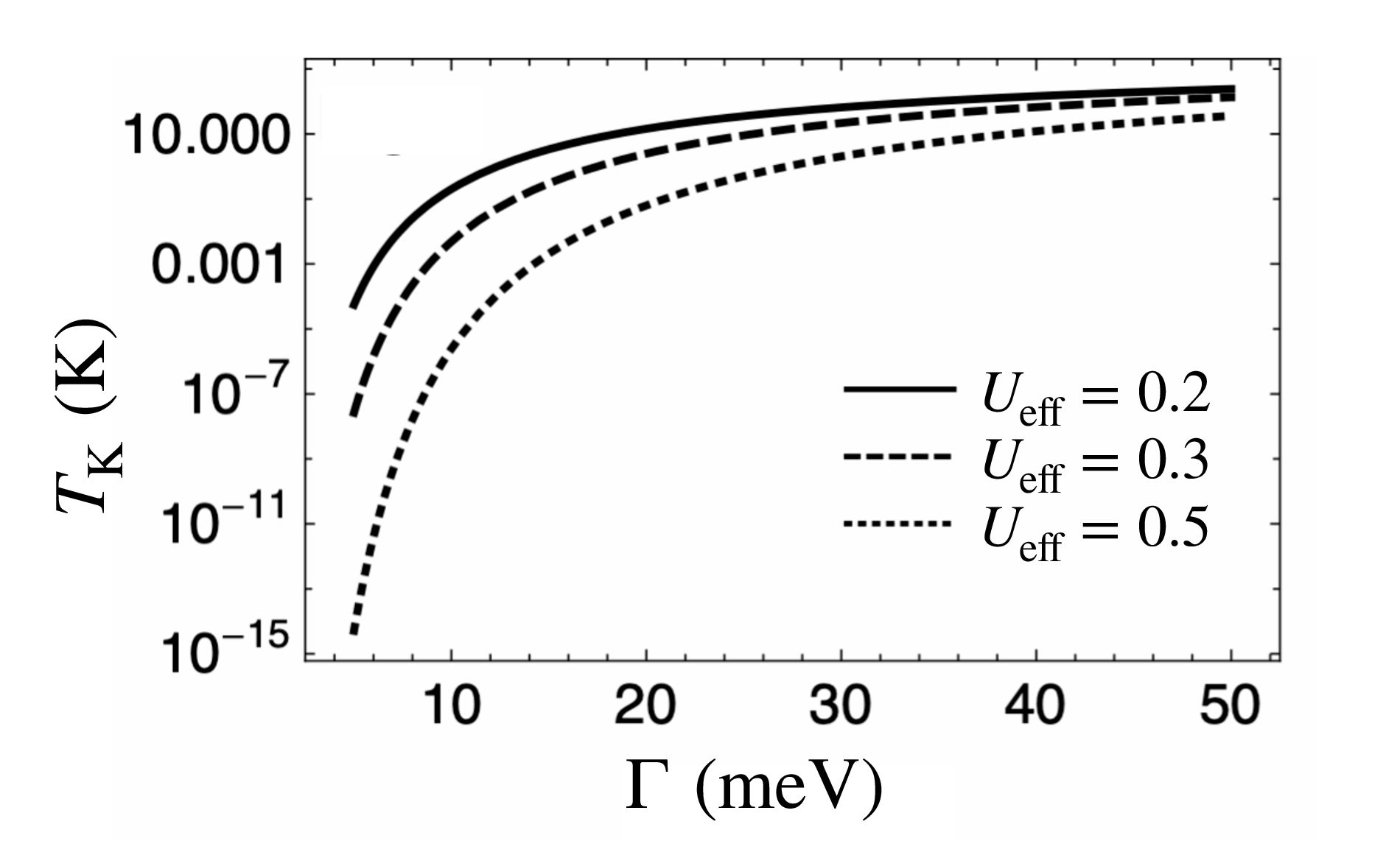}
\caption{
Kondo temperature $T_{\rm K}$ as a function of hybridization strength $\Gamma$ for representative effective Coulomb repulsions $U_{\rm eff}=0.2$, $0.3$, and $0.5$~eV, evaluated using Eq.~(\ref{eq:TK_PH}). 
The strong exponential dependence implies that small variations in adsorption geometry can induce drastic changes in $T_{\rm K}$.
}
\label{fig:TKvsGamma}
\end{figure}

We first evaluate the Kondo temperature expected for a soliton-induced impurity on Au(111). 
In Fig.~\ref{fig:TKvsGamma} we plot $T_{\rm K}$ obtained from Eq.~(\ref{eq:TK_PH}) as a function of $\Gamma$ for representative interaction strengths $U_{\rm eff}=0.2$, $0.3$, and $0.5$~eV. 
The result clearly demonstrates the extreme exponential sensitivity of the Kondo scale to hybridization. 
In the physically relevant range $\Gamma\sim 5$--$50$~meV, $T_{\rm K}$ can vary over many orders of magnitude.

For example, at $U_{\rm eff}=0.3$~eV, changing $\Gamma$ from $10$~meV to $30$~meV increases $T_{\rm K}$ from the mK regime to the experimentally accessible range of tens of Kelvin. 
This implies that relatively small differences in adsorption geometry can determine whether a soliton defect exhibits a measurable Kondo resonance. 
Therefore, the observation of Kondo features is expected to be strongly defect-selective even within nominally identical domain walls.

\subsection{Kondo effect disappearance at topological transition}
\begin{figure}
\centering
\includegraphics[clip,width=8cm]{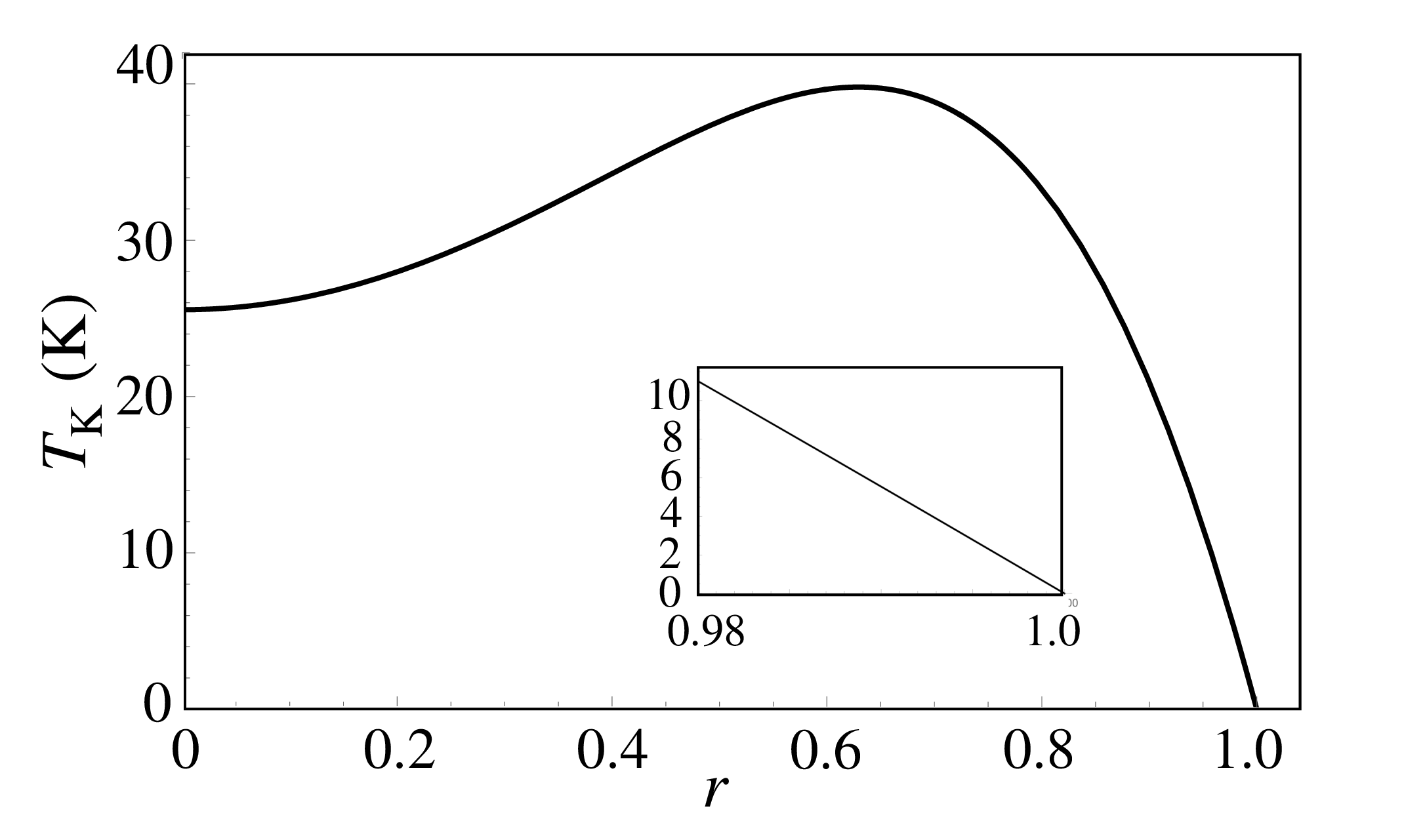}
\caption{
Kondo temperature $T_{\rm K}$ as a function of topological parameter $r=t_1/t_2$ for $U=0.6$ eV, $\Gamma_0=0.005$ eV evaluated using Eq.~\eqref{eq:TK_topo}. 
The inset shows the enlarged view in the vicinity of the topological transition point $r=1$.  
}
\label{fig:TKVSr}
\end{figure}

In this part, we consider the interplay between the topological phase transition and the Kondo effect. 
Substituting Eqs.~\eqref{eq:UeffTopxi} and \eqref{eq:GammaiTop} into Eq.~\eqref{eq:TK_PH}, we obtain 
\begin{equation}
T_{\rm K} \sim \frac{1-r^2}{2}\sqrt{\frac{1}{(1+r^2)}\Gamma_0 U}
\exp\left[-\frac{\pi (1+r^2)U}{8\Gamma_0}\right],   
\label{eq:TK_topo}
\end{equation}
where we approximate the soliton orbital as effectively localized on the defect site $N_{\rm s}$. 
Eq.~\eqref{eq:TK_topo} implies the Kondo temperature approaches zero for $r\to 1$ (See Fig.~\ref{fig:TKVSr}). 
As shown in the inset of Fig.~\ref{fig:TKVSr}, this demonstrates that the many-body scale vanishes linearly with the topological mass parameter $1-r^2$, and hence linearly in $1-r$ close to the transition. 
We note that the linear collapse near $r\to 1$ originates from the prefactor proportional to $1-r^2$, while the exponential factor varies only weakly with $r$ in the physically relevant regime. 
This behavior is natural since the solitonic mode no longer behaves as an impurity for large $\xi$ and completely vanishes at $r=1$, namely $t_1=t_2$. 
It should be noted that the opposite limit $r\to 0$ yields $U_{\rm eff}\to U/2$ and $\Gamma\to \Gamma_0/2$ and thus the impurity Anderson model is achieved. 

\subsection{Height dependence of hybridization on Au(111)}
\label{kondoz}

The exponential sensitivity of $T_{\rm K}$ becomes physically transparent by considering the adsorption-height dependence of the hybridization. 
For molecular adsorbates on Au(111), the tunneling matrix element decays as
\begin{equation}
V(z)\sim V_0 e^{-\kappa z},
\label{eq:Vz}
\end{equation}
yielding
\begin{equation}
\Gamma(z)\propto |V(z)|^2
\sim \Gamma_0 e^{-2\kappa z}.
\label{eq:Gz}
\end{equation}

\begin{figure}[t]
\centering
\includegraphics[clip,width=8cm]{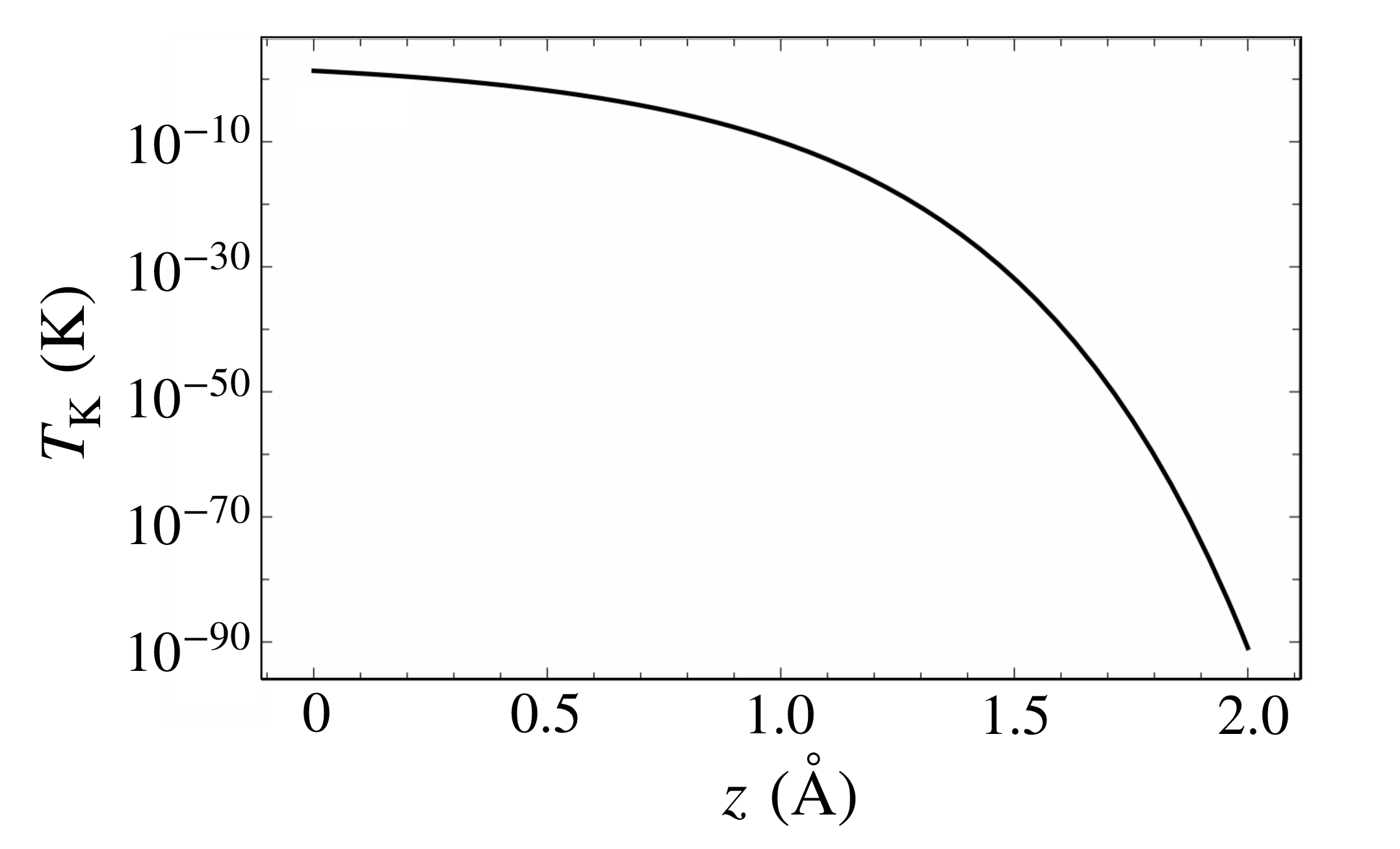}
\caption{
Adsorption-height dependence of the Kondo temperature $T_{\rm K}$ on Au(111), assuming an exponential decay of the hybridization $\Gamma(z)\sim \Gamma_0 e^{-2\kappa z}$. 
A sub-Angstr\"om variation in adsorption height can suppress $T_{\rm K}$ by several orders of magnitude.
}
\label{fig:TKvsZ}
\end{figure}

Combining Eq.~(\ref{eq:Gz}) with Eq.~(\ref{eq:TK_PH}) implies that the Kondo scale is extremely sensitive to adsorption height. 
Figure~\ref{fig:TKvsZ} illustrates this effect, where $T_{\rm K}$ is plotted as a function of a relative height shift $z$ for a representative decay constant $\kappa\sim 1~\text{\AA}^{-1}$. 
A height variation of order $\Delta z\sim $0.5 -- 1$~\text{\AA}$, which is entirely realistic due to local strain or bonding configuration, is sufficient to suppress the Kondo temperature by several orders of magnitude.

This result provides a natural explanation for heterogeneous experimental behavior: soliton-bound states may exist generically at domain walls, but only those defects with sufficiently strong coupling to the Au substrate will display a pronounced Kondo anomaly. 
The result that $T_{\rm K}$ varies by orders of magnitude with sub-Angström changes in adsorption height (as shown in Fig. \ref{fig:TKvsZ} ) indicates that the Kondo effect in these systems is extremely sensitive to the local environment. 
This is consistent with the "Kondo switch" behavior previously reported in magnetic molecules \cite{Zhao2005}, where the Kondo resonance is toggled by subtle changes in chemical bonding with the substrate. 
Crucially, our work extends this concept by demonstrating that while band topology protects the existence of the soliton state, it does not provide a similar "topological protection" for the many-body energy scale itself. Instead, the spatial extension of the topological state dictates the degree of its environmental sensitivity, thereby linking the macroscopic topological phase to the microscopic variability of Kondo screening.

\subsection{Fano lineshape of the Kondo resonance}

\begin{figure}[t]
\centering
\includegraphics[clip, width=8cm]{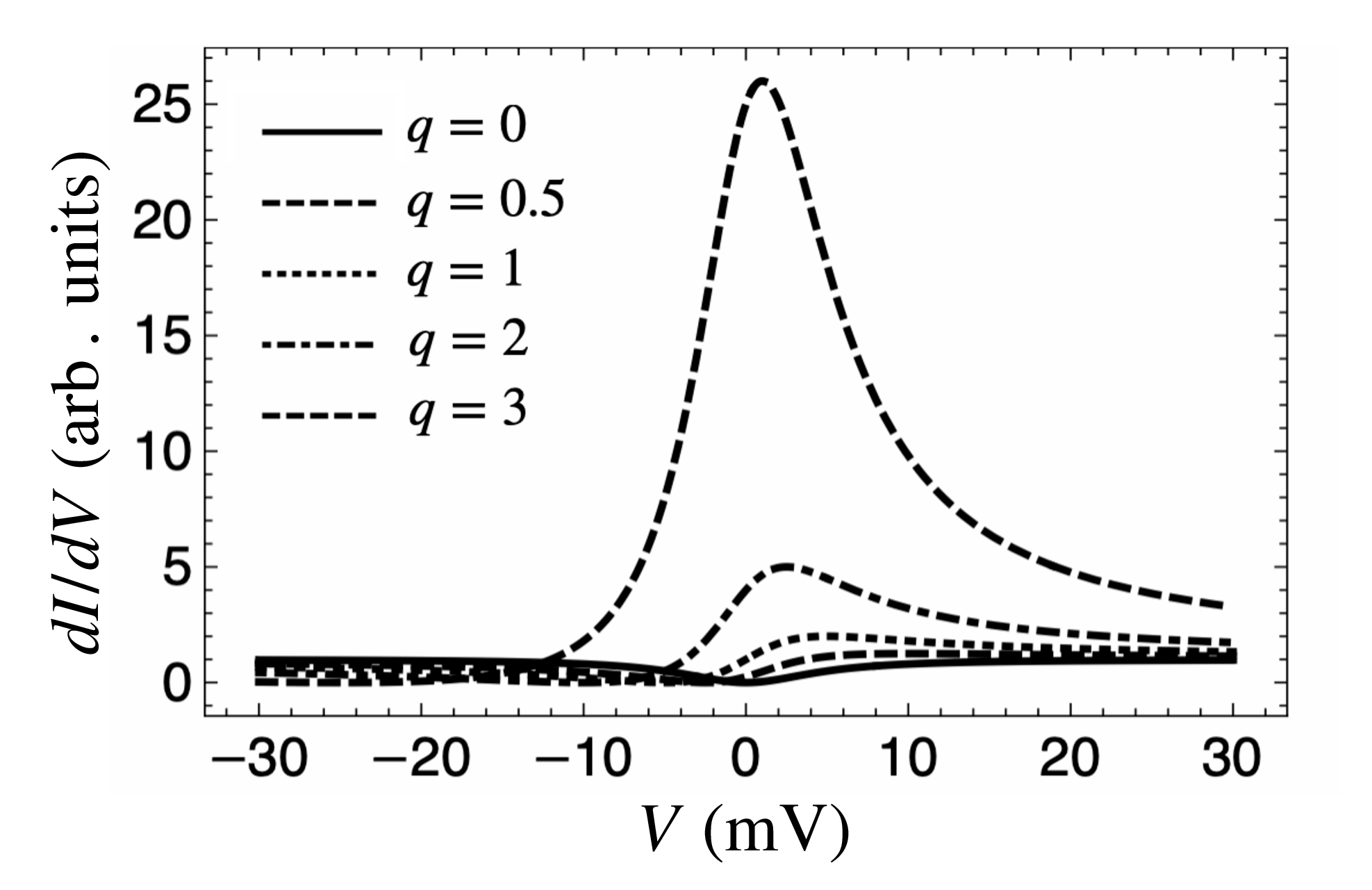}
\caption{
Representative Fano lineshapes for different values of the asymmetry parameter $q$ with $a=1$, $b=0$, $V_0=0$, and $\Gamma_{\rm K}=5$ meV. 
The Kondo resonance can appear as either a peak or a dip depending on tunneling interference.
}
\label{fig:Fano}
\end{figure}

We next consider the expected STS signature. 
The Kondo resonance is generically distorted by interference effects, leading to a Fano profile. 
Figure~\ref{fig:Fano} shows representative spectra generated from Eq.~(\ref{eq:Fano}) for several values of $q$ with $a=1$, $b=0$, $V_0=0$, and $\Gamma_{\rm K}=5$. 
For $q=0$, the feature appears as a dip-like antiresonance, while for $q\gtrsim 1$ it becomes an asymmetric peak. 
In the large-$q$ limit, the resonance approaches a Lorentzian peak.

This implies that the experimentally observed zero-bias feature can appear either as a peak or as a dip, depending on the local tunneling geometry. 
Consequently, a dip-like anomaly at the soliton position is not inconsistent with Kondo physics; rather, it is a natural consequence of Fano interference (For more detail, see Appendix \ref{appB}).

\subsection{Temperature evolution of the zero-bias anomaly}

A central hallmark of the Kondo effect is the suppression and broadening of the zero-bias anomaly with increasing temperature. 
To illustrate this behavior, we adopt a phenomenological temperature-dependent width \cite{Nagaoka2002}
\begin{equation}
\Gamma_{\rm K}(T)=\Gamma_{\rm K}(0)\sqrt{1+(T/T_{\rm K})^2},
\label{eq:GammaKT}
\end{equation}
which captures the crossover from the strong-coupling regime ($T\ll T_{\rm K}$) to the weak-coupling regime ($T\gg T_{\rm K}$).

\begin{figure}[t]
\centering
\includegraphics[clip, width=8cm]{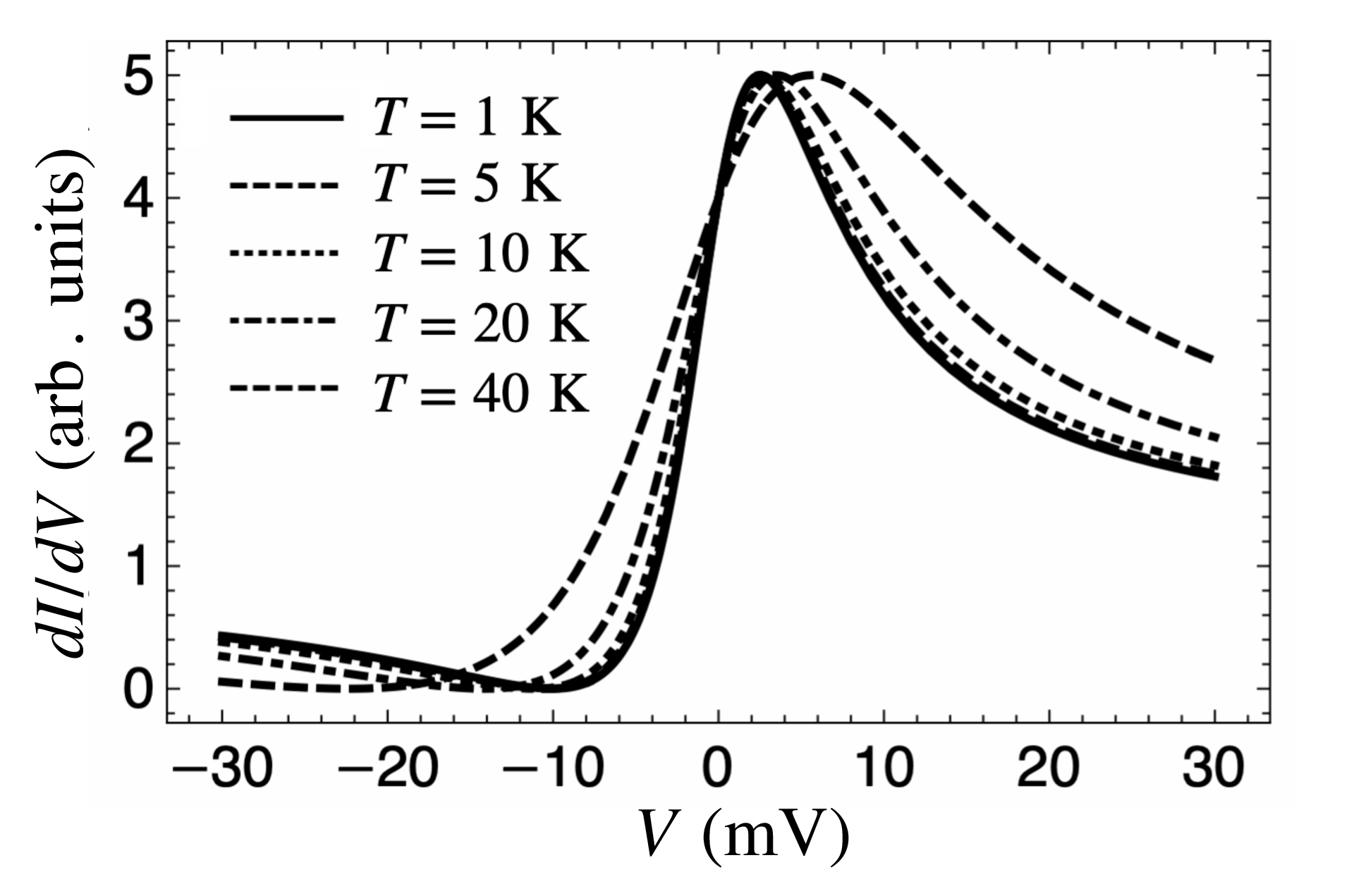}
\caption{
Temperature evolution of the Fano-type Kondo resonance for $q=2$ with $a=1$, $b=0$, $V_0=0$, and $T_{\rm K}=20$ K. 
The zero-bias anomaly broadens and is suppressed as $T$ approaches and exceeds $T_{\rm K}$.
}
\label{fig:Tdep}
\end{figure}

Figure~\ref{fig:Tdep} shows the evolution of the STS spectra for $T_{\rm K}=20$~K with $q=2$. 
The resonance becomes progressively broadened and suppressed as $T$ increases. 
This provides a direct method to extract the Kondo scale experimentally: by fitting measured $dI/dV$ curves to Eq.~(\ref{eq:Fano}) and tracking the fitted width $\Gamma_{\rm K}(T)$, one can obtain a consistent estimate of $T_{\rm K}$.

\if0
\subsection{Magnetic-field splitting}

Another decisive diagnostic is the response to an applied magnetic field. 
The Zeeman energy is approximately
%
\begin{equation}
\Delta E \approx g\mu_B B,
\label{eq:Zeeman}
\end{equation}
%
with $g\approx 2$ and $\mu_B$ the Bohr magneton. 
This yields $\Delta E\approx 0.116~\text{meV/T}$. 
Therefore, at $B=5$--$10$~T the splitting is expected to be $0.6$--$1.2$~meV, which is within the energy resolution of low-temperature STM.

Figure~\ref{fig:Bdep} illustrates the expected field-induced splitting of the zero-bias resonance. 
For small enough $T_{\rm K}$, the splitting should be clearly resolved, whereas for large $T_{\rm K}$ the intrinsic resonance width may exceed $\Delta E$, making the splitting difficult to detect. 
This suggests that weaker-Kondo solitons provide a clearer magnetic-field signature.
\fi

\if0
\begin{figure}[t]
\centering
\includegraphics[clip, width=8cm]{Fig5_B_splitting.eps}
\caption{
Magnetic-field splitting of the zero-bias Kondo resonance. 
The resonance is expected to split when the Zeeman energy becomes comparable to the Kondo scale.
}
\label{fig:Bdep}
\end{figure}
\fi

\subsection{Spatial localization of the Kondo feature}

The soliton orbital is exponentially localized at the domain wall. 
Therefore, the Kondo resonance should also be spatially localized, with its intensity following the soliton wave-function envelope. 
Denoting the soliton wave function by $\psi^{\rm soliton}(x)$ in the continuuum approximation, the local spectral weight at zero bias should approximately scale as
\begin{equation}
\left.\frac{dI}{dV}(x)\right|_{V_{\rm bias}=0}\propto |\psi^{\rm soliton}(x)|^2,
\label{eq:SpatialDecay}
\end{equation}
decaying away from the domain wall on the soliton localization length scale $\xi$. 
This spatial confinement provides an additional strong experimental discriminator, since substrate-induced inhomogeneities would not typically exhibit the same exponential localization tied to the defect.

% ============================================================
\section{Discussion}
% ============================================================

\subsection{Why Kondo physics is not always visible}

As quantified in Section \ref{kondoz}, the extreme sensitivity of $T_{\rm K}$ to the adsorption geometry explains why nominal identity does not guarantee identical spectroscopic signatures.

The above results show that soliton-induced Kondo screening on Au(111) is generically possible but highly configuration dependent. 
Despite the topological robustness of these states, experimental observations often exhibit significant heterogeneity. 
For instance, STS mappings of topological end states in GNRs \cite{Ruffieux2016, Li2020} reveal that the localized signatures can vary from site to site, likely due to local variations in the chemical environment or substrate coupling.

While the existence of a soliton bound state is protected by the topology of the dimerized chain, the Kondo temperature is controlled by the hybridization strength $\Gamma$, which depends sensitively on adsorption geometry. 
This provides an important conceptual separation: the presence of a soliton bound state is primarily a single-particle topological effect, whereas the emergence of a measurable Kondo anomaly is a many-body phenomenon controlled by the strength of coupling to the metallic bath. 
Thus, failure to observe a Kondo resonance at a particular defect does not imply the absence of a soliton state; it may simply indicate that $T_{\rm K}$ lies below the experimental temperature window.

\subsection{Choice of values of $U_{\rm eff}$}
The effective on-site Coulomb interaction, $U_{\rm eff}$, is a key parameter determining the many-body ground state of the soliton. While the bare Coulomb repulsion for isolated organic molecules is typically in the range of several eV, it is significantly renormalized upon adsorption on a metallic substrate due to the image-charge effect and dynamical screening by the conduction electrons \cite{Neaton2006}. Following previous studies on topological molecular chains and graphene nanoribbons on Au(111) \cite{Groning2018, Rizzo2018}, we adopt $U_{\rm eff}$ values between 0.2 and 0.5 eV. This range is consistent with the experimental observation of Kondo scales in similar molecular architectures. Furthermore, we emphasize that the primary conclusion of this work—the exponential sensitivity of $T_{\rm K}$ to the hybridization —remains qualitatively robust across this range of $U_{\rm eff}$.

\subsection{Distinguishing Kondo resonance from trivial in-gap states}

Experimentally, an in-gap resonance at a defect can arise from either (i) a noninteracting bound state, (ii) a resonance broadened by substrate coupling, or (iii) a genuine Kondo resonance. 
Our analysis suggests that Kondo physics can be unambiguously identified by a combination of signatures:
(i) a Fano lineshape centered near zero bias,
(ii) temperature scaling governed by a single scale $T_{\rm K}$, 
and (iii) spatial localization consistent with the soliton envelope.

A purely single-particle resonance would not show the universal temperature scaling associated with Kondo screening. 

We must distinguish the soliton-induced Kondo resonance from vibration-induced features (Inelastic Electron Tunneling Spectroscopy, IETS) that often appear in the same energy range ($10 - 50$ meV).

Two primary criteria can be used:
\begin{itemize}
\item{}Temperature Dependence: The Kondo resonance follows a universal scaling law, significantly broadening and decreasing in amplitude as $T$ approaches $T_{\rm K}$. 
In contrast, IETS features remain relatively sharp, with their width primarily determined by the thermal broadening of the tip's Fermi surface. 
\item{}Magnetic Field Response: Under a high magnetic field, the Kondo resonance undergoes a characteristic Zeeman splitting or suppression. Conversely, molecular vibrations are typically insensitive to magnetic fields of several Tesla. These distinct responses allow for a clear identification of the many-body origin of the observed zero-bias anomaly.
\end{itemize}

\subsection{Role of particle-hole asymmetry}

Although our quantitative plots were obtained at particle-hole symmetry, realistic soliton orbitals may deviate from $\epsilon_d=-U_{\rm eff}/2$ due to charge transfer to the substrate. 
Such asymmetry modifies the effective exchange coupling and shifts the resonance center $V_0$ (See Appendix \ref{appA}). 
However, the essential qualitative feature remains: the Kondo temperature retains its exponential dependence on $\Gamma$, and the STS signal remains describable by a Fano form with a temperature-dependent width. 

Therefore, the predictions made here are robust against moderate particle-hole asymmetry.

\subsection{Interference between modes in the domain wall and the edge}
In finite chains, the proximity of the domain wall to the chain ends can lead to hybridization between the soliton state and the topological edge states (Fig.~\ref{fig:ChemStructure} e). 
The E=0 degeneracy splits when the distance between the domain wall and the end is shorter than the localization length, and the two states overlap. 
This overlap significantly reduces the local density of states at the Fermi level, potentially suppressing the Kondo effect or shifting the resonance energy away from zero bias. 
For the experimental identification of pure soliton-induced Kondo physics, it is therefore crucial to investigate domain walls located at least several unit cells away from the chain ends to ensure the soliton's topological isolation. 
In addition, this interference may induce the RKKY type spin-spin interaction, and that compete with the Kondo effect. 

Beyond the experimental considerations discussed above, it is instructive to reconsider our results from a more general theoretical perspective. 
In the SSH model, the bulk energy gap is directly tied to the Dirac mass controlling the localization length of the domain-wall state. 
While the preceding analysis focused on adsorption geometry and hybridization strength, the underlying mechanism can be reinterpreted in terms of how the topological mass governs the effective interaction scale of the localized mode. 

\subsection{Recast in terms of the bulk gap and topological mass}

Beyond the microscopic parametrization adopted in the previous sections, the present results admit a more transparent physical interpretation in terms of the bulk gap and the associated Dirac mass of the SSH chain. 
In the continuum limit, the dimerization strength controls the bulk energy gap $\Delta$, which in turn determines the inverse localization length of the domain-wall mode. 
The parameter $r$ introduced above is directly related to this gap as
\begin{equation}
1-r^2 = 1-\left|\frac{t_1}{t_2}\right|
= \frac{|t_2|-|t_1|}{|t_2|}
\propto \Delta .
\label{randdelta}
\end{equation}
Since the bulk gap is proportional to the topological mass $m_{\rm top}$ of the low-energy Dirac description,
\begin{equation}
m_{\rm top} \propto \Delta ,
\label{mtopanddelta}
\end{equation}
the collapse of the Kondo scale near $r \to 1$ can be equivalently interpreted as the vanishing of the Dirac mass.
Using these relations, Eq.~\eqref{eq:TK_topo} can be rewritten in the vicinity of the transition as
\begin{align}
T_{\rm K}
&\sim A \Delta \sqrt{\frac{1}{2}\Gamma_0 U}
\exp\left[-\frac{\pi U}{16\Gamma_0}\right]\nonumber\\
&\sim B m_{\rm top} \sqrt{\frac{1}{2}\Gamma_0 U}
\exp\left[-\frac{\pi U}{16\Gamma_0}\right],
\end{align}
where $A$ and $B$ collect nonuniversal prefactors.

This expression makes explicit that the emergent many-body scale is linearly tied to the bulk mass scale. 
Importantly, the disappearance of $T_{\rm K}$ at the topological transition is therefore not an accidental lattice-specific feature, but a direct consequence of the vanishing Dirac mass that controls the spatial confinement of the soliton mode. 
In this sense, the topological mass does not merely determine the existence of the bound state; it quantitatively governs the strength of correlation effects associated with it.

The robustness of this mass-controlled scaling within the SSH realization suggests that similar mechanisms may arise in other domain-wall systems where localized modes inherit their confinement length from a bulk mass parameter.

The linear dependence on $m_{\rm top}$ has a simple physical origin. 
In the continuum description, the topological mass sets the inverse localization length of the soliton mode, $\xi^{-1} \propto m_{\rm top}$. 
As the transition is approached, the bound state delocalizes and its weight at the impurity site decreases proportionally to $m_{\rm top}$. 
Since the effective hybridization entering the Kondo scale is controlled by this local weight, $\Gamma_{\rm eff} \propto m_{\rm top}$, the resulting Kondo temperature inherits the same linear prefactor. 
The collapse of $T_{\rm K}$ at the topological transition is therefore a direct manifestation of critical delocalization of the bound state.

\subsection{Limitations and outlook}
While more sophisticated numerical methods, such as numerical renormalization group (NRG) calculations, may refine quantitative prefactors, the exponential scaling structure and the topology-controlled collapse of the Kondo scale follow directly from the analytical mapping and are therefore robust. 
In particular, the linear collapse of the Kondo temperature at the topological transition and its exponential sensitivity to hybridization arise directly from the analytical structure of the effective model and are therefore robust features independent of specific numerical details.

% ============================================================
\section{Experimental proposal: How to observe soliton-Kondo on Au(111)}
% ============================================================

A promising platform is a dimerized molecular chain assembled on Au(111) via on-surface synthesis or manipulation. 
The dimerization pattern can be stabilized by molecular bonding geometry. 
Domain walls can be introduced either by deliberate termination, by missing-molecule defects, or by constructing a junction between two dimerization domains.

To detect Kondo physics, one should perform $dI/dV$ spectroscopy at the soliton center and away from it. 
A pronounced zero-bias anomaly should appear only near the domain wall. 
The spectrum should then be fitted to the Fano form Eq.~(\ref{eq:Fano}). 
The extracted width $\Gamma_{\rm K}$ yields an estimate of the Kondo temperature via $k_BT_{\rm K}\sim \Gamma_{\rm K}$. 
The key test is consistency: $T_{\rm K}$ extracted from the width should match the temperature at which the anomaly disappears.

Temperature-dependent STS measurements should be performed from the base temperature (typically $T\sim 1$--$5$~K) up to tens of Kelvin. 
The resonance is expected to broaden and weaken with increasing $T$. 
A consistent fit to Eq.~(\ref{eq:GammaKT}) should yield a characteristic $T_{\rm K}$.

Finally, a magnetic field applied perpendicular to the Au(111) surface should induce Zeeman splitting. 
For $B=5$--$10$~T, the splitting energy is expected to be $0.6$--$1.2$~meV for $g=2$. 
Therefore, high-resolution STS with meV-scale resolution should resolve two peaks when $T_{\rm K}$ is not too large. 
Spatially resolved $dI/dV(x,V)$ maps should further demonstrate exponential localization at the domain wall.

% ============================================================
\section{Conclusion}
% ============================================================

We have developed a theoretical framework demonstrating that topology can directly regulate a many-body energy scale in molecular systems. 
By mapping the SSH-Hubbard model onto an effective Anderson impurity description of the soliton orbital hybridized with a metallic substrate, we analytically established that the Kondo temperature is governed by the topological dimerization parameter and vanishes linearly at the topological transition. 
This result reveals a direct proportionality between topological mass and many-body screening scale.
This establishes a minimal analytical mechanism by which a bulk Dirac mass controls an emergent correlation scale.

In addition to this topological control, the Kondo temperature retains its exponential sensitivity to the hybridization strength $\Gamma$, implying that sub-Angström variations in adsorption geometry can produce order-of-magnitude changes in the observable Kondo signal. 
Such combined topological and geometrical regulation naturally explains the apparent heterogeneity of Kondo signatures in experiments. 
We further predicted characteristic Fano lineshapes and temperature evolution in scanning tunneling spectroscopy, providing concrete experimental criteria for identifying topology-controlled Kondo screening.

In summary, our work establishes topological domain walls in dimerized molecular chains on Au(111) as a platform where a Dirac mass term directly governs a many-body energy scale. 
This study highlights a general mechanism by which topology controls correlation effects and offers a conceptual and quantitative framework for engineering many-body quantum states in molecular architectures.
The linear collapse of the Kondo scale at the topological transition provides a direct experimental route to detect topology through many-body spectroscopy.
More broadly, our work demonstrates that topology does not merely protect states, but can quantitatively determine emergent many-body energy scales, opening a route toward topologically engineered correlated quantum matter.

\section*{ACKNOWLEDGEMENTS}
The work of R.~Y.~is supported by JSPS KAKENHI No.~JP25K07156. 
A related study has appeared in Ref.~\cite{Aligia2025}, which addresses complementary aspects of the problem. 
In contrast to Ref.~\cite{Aligia2025}, which focuses on numerical modeling, we provide an analytical scaling theory linking topology to the many-body scale.

% ============================================================
\appendix
\section{Schrieffer--Wolff transformation}
\label{appA}
% ============================================================

In the local-moment regime of the Anderson impurity model, charge fluctuations to the empty and doubly occupied impurity states are virtual processes. 
To second order in the hybridization, one may integrate out these charge fluctuations and obtain an effective Kondo exchange Hamiltonian.

Starting from
\begin{equation}
H = H_{\rm bath} + H_d + H_{\rm hyb},
\end{equation}
with
\begin{equation}
H_d = \sum_\sigma \epsilon_d d^\dagger_\sigma d_\sigma
+U_{\rm eff}n_{d\uparrow}n_{d\downarrow},
\end{equation}
and
\begin{equation}
H_{\rm hyb}=\sum_{k,\sigma}(V_k c^\dagger_{k\sigma}d_\sigma+{\rm h.c.}),
\end{equation}
the Schrieffer--Wolff transformation yields
\begin{equation}
H_{\rm eff}=H_{\rm bath}+J\bm{S}\cdot \bm{s}(0)
+W\sum_\sigma c^\dagger_{0\sigma}c_{0\sigma},
\end{equation}
where $W$ is a potential scattering term. 
The exchange coupling is
\begin{equation}
J = 2|V|^2
\left(
\frac{1}{|\epsilon_d|}
+
\frac{1}{\epsilon_d+U_{\rm eff}}
\right),
\end{equation}
and the potential scattering is
\begin{equation}
W = |V|^2
\left(
\frac{1}{|\epsilon_d|}
-
\frac{1}{\epsilon_d+U_{\rm eff}}
\right).
\end{equation}

At particle-hole symmetry $\epsilon_d=-U_{\rm eff}/2$, the potential scattering vanishes $W=0$, and one obtains a pure Kondo model.

% ============================================================
\section{Connection to Fano fitting}
\label{appB}
% ============================================================

The Fano expression used in Eq.~(\ref{eq:Fano}) arises from interference between tunneling amplitudes into the localized impurity orbital and the substrate continuum. 
In STM experiments, the tunneling current is proportional to the imaginary part of the local Green's function at the tip position. 
When both impurity and substrate channels contribute, the resulting conductance naturally acquires the Fano form.

Thus, the extracted $q$ parameter encodes the ratio of tunneling matrix elements into the soliton orbital and into the Au substrate. 
Variations in $q$ between different soliton defects are expected due to local geometry and do not contradict a universal Kondo interpretation.

The Fano parameter $q$ represents the ratio of tunneling amplitudes into the discrete soliton state versus the continuum of the metal substrate. 
Specifically, $q=\frac{\langle f|H|i\rangle }{\pi V \langle f|H|k\rangle }$, where $|i\rangle $ is the soliton state and $|k\rangle $ is the substrate continuum.
When the STS tip is positioned directly above the soliton center at the domain wall, the direct tunneling into the molecular orbital is maximized, typically resulting in a large $|q|$ (peak-like feature). 
As the tip moves laterally away from the defect center, the coupling to the localized soliton wave function decays faster than the coupling to the metallic surface, leading to a decrease in $|q|$ and a transition toward a dip-like structure ($q\approx 0$). 
Furthermore, a larger adsorption height $z$ generally decreases the overall hybridization $\Gamma$, but also modifies the ratio of tunneling paths, often leading to a more symmetric dip if the orbital symmetry and the tip-sample distance favor interference with the substrate's surface state.

% ============================================================

\end{document}